\documentclass[aps,prb,preprint,a4paper]{revtex4}
\usepackage{amsmath}
\usepackage{amsfonts}
\usepackage{graphicx}
\usepackage{bbm}

\newcommand{\remark}[1]{}

\begin{document}

\preprint{ITP-UH-03/06}

\title{Anderson-like impurity in the one-dimensional $t$--$J$ model: formation
of local states and magnetic behaviour}

\author{Holger Frahm}
\author{Guillaume Palacios}
\affiliation{Institut f\"ur Theoretische Physik, Universit\"at Hannover,
             Appelstr.\ 2, 30167 Hannover, Germany}

\date{March 08, 2006}

\begin{abstract}
We consider an integrable model describing an Anderson-like impurity coupled
to an open $t$--$J$ chain.  Both the hybridization (i.e.\ its coupling to bulk
chain) and the local spectrum can be controlled without breaking the
integrability of the model.  As the hybridization is varied, holon and spinon
bound states appear in the many body ground state.  Based on the exact
solution we study the state of the impurity and its contribution to
thermodynamic quantities as a function of an applied magnetic field.  Kondo
behaviour in the magnetic response of the impurity can be observed provided
that its parameters have been adjusted properly to the energy scales of the
holon and spinon excitations of the one-dimensional bulk.
\end{abstract}
\maketitle

\section{Introduction}
The possibility of controlled embedding of quantum impurities, i.e.\ local
scatterers with internal degrees of freedom, into nanofrabricated devices has
led to new manifestations of Kondo physics, e.g.\ in quantum dot systems or
atoms deposited on metallic surfaces\cite{NanoKondo}.  In these
systems many parameters of the impurity, such as its internal spectrum, its
coupling to the metallic environment and the properties of the latter can be
tuned within the experiment.  Investigating the effects of these parameters on
observable quantities leads to various new questions: originally, most of the
theoretical work on the quantum impurity problems has neglected the effect of
electronic correlations in the host system\cite{Hewson}.  If the latter is
one-dimensional, however, any interaction leads to non-Fermi liquid behaviour.
The low energy regime is then described by a Tomonaga Luttinger liquid (TLL)
characterized by continuously varying exponents of its ground state
correlation functions\cite{GoNeTs:boson}.  As a consequence, the local
density of states vanishes as a power law in such a system.  Therefore it is
to be expected that the critical properties of the impurity will be strongly
affected.  This problem of a quantum impurity coupled to a TLL has been
subject of intense studies in recent years -- both analytically based on field
theoretical methods and numerically.  These studies indicate that depending on
the coupling parameters different non-trivial fixed points at intermediate or
strong coupling can be realized which determine the observables accessible to
experiments (see e.g.\ Refs.~\onlinecite{LeTo92,FuNa94,FrJo95,EgKo98,Furu05}).

This picture of the low energy behaviour of these systems has to be supported
by methods which do not rely on the analysis at weak coupling, e.g.\ exact
solutions as for the Kondo and Anderson impurity problem in a Fermi
liquid\cite{KondoBA}.
For results which cover the full range of experimentally available parameters
specific realizations of quantum impurity models have to be studied.  An
approach which allows to make contact to the universal low energy behaviour
identified using the methods mentioned above is based on integrable lattice
models.  Starting, e.g., from the Bethe ansatz solvable supersymmetric
$t$--$J$ model for electrons on a one-dimensional lattice or variations
thereof one can consider different representations of the graded Lie algebra
$gl(2|1)$ for the spectrum of electronic states on bulk and impurity sites
without destroying integrability\cite{BEFimp,ScZv9x,Foer99a,BoKS05}.  The
physical properties of these impurities can be tuned by variation of the
representation (which may depend on a continuous parameter for $gl(2|1)$) and
by a shift in the spectral parameter which directly enters the coupling
between the impurity and the host system.

The resulting Hamiltonian of models constructed along these lines takes a
particularly simple form when such impurities are combined with open
boundaries\cite{FrZv97b,tJKondo,Zhou99}.  In this paper we shall consider the
system introduced in Ref.~\onlinecite{BeFr99a}, based on the supersymmetric
$t$--$J$ chain with open boundary conditions.  The Hamiltonian of this model
is
\begin{eqnarray}
  {\cal H}&=& -{\cal P}\left(\sum_{j=2}^{L}
        \sum_\sigma c^\dagger_{j,\sigma}c_{j+1,\sigma}
        +c^\dagger_{j+1,\sigma}c_{j,\sigma}\right){\cal P}
\nonumber\\
  &&+2\sum_{j=2}^{L} \left[ {\vec S_j}{\vec S_{j+1}}
        -\frac{n_jn_{j+1}}{4} +{1 \over 2} (n_j+n_{j+1}) \right]  
        -HS^z -\mu N+{\cal H}_{b}\ ,
\label{ham}
\end{eqnarray}
where ${\cal P}$ projects out double occupancies on the bulk sites ($j=2$ to
$L+1$) and $\vec{S_j}=\sum_{\alpha\beta} c^\dagger_{j,\alpha}
\vec{\sigma}_{\alpha\beta} c_{j,\beta}$, $n_j = \sum_\sigma
c^\dagger_{j,\sigma}c_{j,\sigma}$ are the electronic spin and number operators
on site $j$.  The magnetization of the system is controlled by the field $H$
and we consider the system at fixed hole concentration $\delta=1-\sum_j n_j/L$
(alternatively one may use a grand canonical approach to control $\delta$ by
variation of the chemical potential $\mu$).
The impurity is added at the boundary (site $1$) of this chain, its internal
spectrum and coupling to the bulk is determined by
\begin{equation}
{\cal H}_b = \frac{4}{4t^2 + (\alpha +2)^2} \left[\alpha n_2 + 
2{\vec S_1}{\vec S_2} -\frac{n_1 n_2}{2}
+n_1 +n_2 -\sum_{\sigma}(Q_{2,\sigma}^{\dagger}Q_{1,\sigma} + 
Q_{1,\sigma}^{\dagger}Q_{2,\sigma})
\right]
\label{hamb}
\end{equation}
where $Q_{2,\sigma}=|0\rangle\langle\sigma|_2$ and $Q_{1,\sigma} =\sqrt{\alpha
+1}|0\rangle\langle\sigma|_1 -2\sigma\sqrt{\alpha}\
|\bar{\sigma}\rangle\langle 2|_1$ are generalized electron annihilation
operators for site $1$ and $2$.  The parameter $\alpha>0$ labels the
four-dimensional representation of $gl(2|1)$ used in the construction of
(\ref{hamb}) and controls the internal spectrum of the impurity.
The terms in (\ref{hamb}) describe exchange and Coulomb interaction between
the electrons on site $1$ and those in the chain as well as a term allowing
for the hopping of electrons between the bulk and the impurity.  Comparing
this model with that of the single-impurity Anderson model, $V_0 \equiv
\frac{4}{4t^2 + (\alpha +2)^2}$ can be identified with a hybridization
coupling.  Since we consider an open chain, the parameter $t$ can be either
real positive or purely imaginary, thereby allowing to cover the entire range
$-\infty<V_0<\infty$ for this coupling between the bulk and the impurity site.
Note that additional parameters can be introduced into the model by adding
static boundary fields\cite{BeFr99a}.  For generic parameters, however, the
$1/L$ contributions of the impurity, boundary and boundary fields to the
thermodynamic properties are additive in the integrable model.  Since the
$t$--$J$ model with open boundaries and boundary fields has already been
studied in great detail\cite{Essl96} we restrict our analysis to the impurity
contributions to the magnetization and the susceptibilities.  These are
functions of the bulk density of holes $\delta$, the magnetic field $H$ and
the parameters controlling the impurity, i.e.\ $\alpha$ and $V_0$.
Furthermore we consider only the ground state properties of the system to
avoid the subtleties arising in the analysis of systems with open boundaries
at finite temperatures\cite{GoBF05}.

In the following section we identify the configuration of the impurity in the
ground state of the many particle system as a function of these parameters.
Then we study the magnetic susceptibility and occupation of the impurity at
zero magnetic field where we derive analytic expressions near half filling
($\delta=0$) which are complemented by numerical results for general
$\delta$.  Based on these results we identify the relevant energy scales in
the  system which allow for a quantitative study of the properties of the
system in an external magnetic field.


\section{Formation of Bound states}%

\subsection{Bethe Ansatz Equations}
The spectrum of the model (\ref{ham}) has already been obtained by means 
of the algebraic Bethe ansatz technique\cite{BeFr99a}. 
An eigenstate with $N_e = N_{\uparrow} + N_{\downarrow}$ electrons 
is characterized by the roots of the Bethe ansatz equations (BAE):
\begin{equation}
\begin{aligned}
& \left(e_1(\lambda_j)\right)^{2L} =\prod_{k\neq j}^{M_s}
        e_2(\lambda_j-\lambda_k) e_2(\lambda_j+\lambda_k) 
        \prod_{\beta=1}^{M_c} e_{-1}(\lambda_j-\vartheta_\beta)
        e_{-1}(\lambda_j+\vartheta_\beta)\,, \quad j=1,\ldots M_s,\\
 & e_{\alpha}(\vartheta_\gamma+t) e_{\alpha}(\vartheta_\gamma-t)  =
        \prod_{k=1}^{M_s}e_{-1}(\vartheta_\gamma-\lambda_k)
     e_{-1}(\vartheta_\gamma+\lambda_k)\,,\quad \gamma=1,\ldots M_c.
\label{bae0}
\end{aligned}
\end{equation}
Here $e_n(x)=\frac{x+{in}/{2}}{x-{in}/{2}}$.  The $M_s=L+1-N_\uparrow$ spin
rapidities $\lambda_j$ parameterize magnetic excitations starting from the
completely filled state with maximum polarization (1 electron per bulk site, 2
electrons on the impurity site with $N_{\uparrow} = L + 1$, $N_{\downarrow}
=1$) and the $M_c=L+2-N_e$ charge rapidities $\vartheta_\gamma$ describe holes
added to this state.  The phases $e_\alpha$ in the second set of these
equations are due to the presence of the impurity (\ref{hamb}), similarly as
in Ref.~\onlinecite{BEFimp}.  The energy of the corresponding Bethe state is
then given by the expression
\begin{equation}
  E=V_0 (\alpha +2) +2(L-1)-
  \sum_{j=1}^{M_s}\frac{1}{\lambda_j^2+\frac{1}{4}}
  +\left(\mu-\frac{H}{2}\right)M_c + H M_s - \mu (L+2) - \frac{H}{2}L.
\label{Egs}
\end{equation}

\remark{\begin{equation}
E_b = \frac{\alpha+2}{t^2+(\alpha/2+1)^2}.
\label{Eb}
\end{equation}}

\begin{figure}[ht]
  \includegraphics[width=0.7\textwidth]{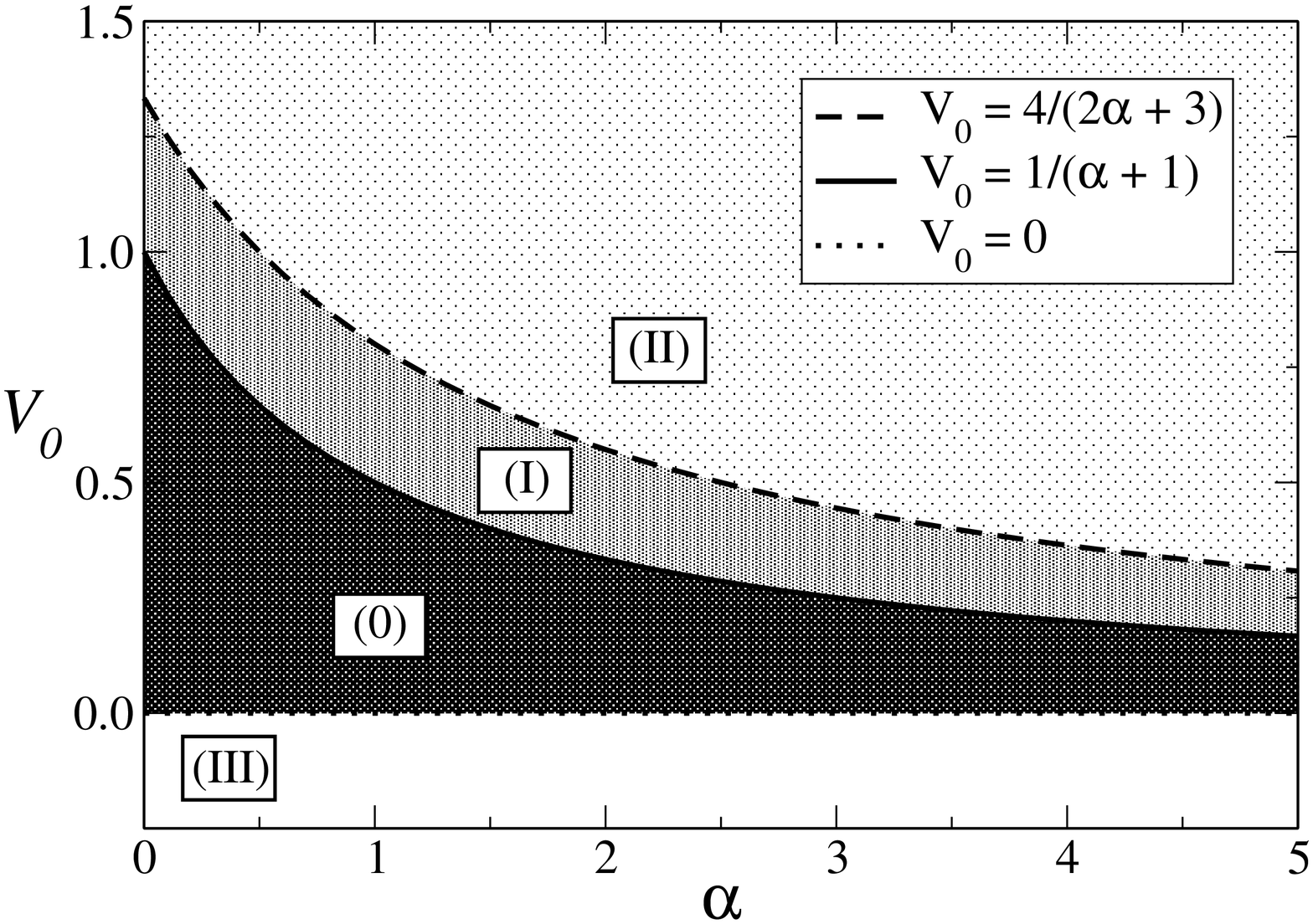}
  \caption{
    \label{V0aplane}
    Spectrum of impurity bound states in the $V_0$--$\alpha$ parameter space.}
\end{figure}

The ground state and the low-energy magnetic and charged excitations 
(spinons and holons) of the $t$--$J$ model with open boundary conditions 
without boundary fields or an impurity - similar to the model with 
periodic boundary conditions - are described by positive rapidities
$\{\lambda_j ,\vartheta_\gamma\}$ solving the BAE (\ref{bae0}).
Sufficiently strong boundary magnetic fields or potentials lead to the
formation of boundary bound states in the spectrum of the system.  In terms of
the many-particle Bethe states this is reflected by the appearance of
isolated, purely imaginary roots to the BAE\cite{SkSa95,BeFr97,TsYa97}.  Note
that a necessary condition for such roots is the existence of singularities in
the boundary phase shifts to compensate divergencies appearing in the
scattering phases for imaginary rapidities.
A similar mechanism for the creation of bound states has been found to exist
in integrable impurity problems\cite{KondoBound}, although it has not been
studied systematically so far.  For the present model with the impurity
described by Eq.~(\ref{hamb}) bound states appear for a sufficiently strong
hybridization $V_0$.  This regime is reached by choosing $t=i\tau$ purely
imaginary ($\tau >0$ without loss of generality).  In the presence of a
particular solution with an imaginary root $\mathrm{Im}(\vartheta_{\gamma_0})
> 0$, the r.h.s. of the second set of equations in (\ref{bae0}) vanishes in
the thermodynamic limit ($M_s \to \infty$ with $M_{s,c}/L$ kept fixed).
Therefore, $\vartheta_\gamma$ has to be exponentially close to a zero of
$e_{\alpha}(\vartheta_\gamma+t) e_{\alpha}(\vartheta_\gamma-t)=
e_{\alpha+2\tau}(\vartheta_\gamma) e_{\alpha-2\tau}(\vartheta_\gamma)$,
i.e. $\vartheta_{\gamma_0} = -i (\alpha/2 \pm \tau)$.  This bound state
solution appears for $\tau > \alpha/2 \equiv \tau_0$. Increasing $\tau$
further eventually leads to the appearance of additional imaginary roots in
the ground state configuration. The analysis of this sequence of bound states
suggests a division of the $V_0$--$\alpha$ parameter space of the impurity
into four regions labeled by an index (R) counting the number of bound state
solutions present in the ground state of the system (see Fig. \ref{V0aplane}).
$R$ runs from 0 (the region described by the BAE (\ref{bae0})) to III (a
region with three bound states).
\begin{itemize}
\item[{\bf (I)}:]{$\tau_0 <\tau<(\alpha + 1)/2 \equiv\tau_1$}

As seen above, a complex root $\vartheta_{M_c}=i(\tau-\alpha/2)$ appears in
the set of the charge rapidities $\{ \vartheta_\gamma\}$ for $\tau > \alpha
/2$.  Explicitly taking into account this root, describing a bound state in
the holon sector, and rearranging the equations we end up with a set of
modified BAE:
 \begin{equation}
 \begin{aligned}
 e_1^{2L}(\lambda_j) =& e_{-(2\tau +1 -\alpha)}(\lambda_j)
                         e_{2\tau-1-\alpha}(\lambda_j)\,
        \prod_{k\neq j}^{M_s}
        e_2(\lambda_j-\lambda_k) e_2(\lambda_j+\lambda_k) \\
  &
        \times\prod_{\beta=1}^{M_c -1} e_{-1}(\lambda_j-\vartheta_\beta)
        e_{-1}(\lambda_j+\vartheta_\beta)\,,\quad j=1,\ldots,M_s,\\
 e_{\alpha+2\tau}(\vartheta_\gamma) e_{\alpha-2\tau}(\vartheta_\gamma) =&
        \prod_{k=1}^{M_s}e_{-1}(\vartheta_\gamma-\lambda_k)
                         e_{-1}(\vartheta_\gamma+\lambda_k)\,,\quad
   \gamma=1,\ldots,M_c-1\ .
\label{bae1}
\end{aligned}
\end{equation}

\item[{\bf (II)}:]{$\tau_1 <\tau<\alpha/2 + 1\equiv\tau_2$}

Increasing $\tau$ even further, we will generate a new bound state, this time,
in the spin sector. We associate to this bound state, the complex spin
rapidity $\lambda_{M_s}=i(\tau-(\alpha +1)/2)$.  Once again, the BAE have to
be changed in consequence:
\begin{equation}
\begin{aligned}
 e_1^{2L}(\lambda_j) =& e_{-(2\tau -\alpha -3)}(\lambda_j)
                            e_{2\tau-1-\alpha}(\lambda_j)
        \prod_{k\neq j}^{M_s -1}
        e_2(\lambda_j-\lambda_k) e_2(\lambda_j+\lambda_k)\\ 
        & \times\prod_{\beta=1}^{M_c -1} e_{-1}(\lambda_j-\vartheta_\beta)
        e_{-1}(\lambda_j+\vartheta_\beta)\,,
  \quad j=1,\ldots,M_s-1,\\
 e_{\alpha+2\tau}(\vartheta_\gamma)  =&
         e_{2\tau-2-\alpha}(\vartheta_\gamma)
        \prod_{k=1}^{M_s -1}e_{-1}(\vartheta_\gamma-\lambda_k)
                e_{-1}(\vartheta_\gamma+\lambda_k)\,,
      \quad\gamma=1,\ldots,M_c-1.
\label{bae2}
\end{aligned}
\end{equation}
From (\ref{Egs}) we see that the contribution of this 'spinon bound state' to
the energy of this Bethe state will be
\begin{equation}
  E_{M_s} = [(\tau - \alpha/2)(1-\tau + \alpha/2)]^{-1}\ .
\label{ebound_s}
\end{equation}

\item[{\bf (III)}:]{$\tau>\tau_2$}

For $\tau > \alpha/2 + 1$, a third bound state is created. Another purely
imaginary root, $\vartheta_{M_c -1}=i(\tau-\alpha/2 -1)$, will coexist 
with a set a real rapidities in the charge sector and with the root of 
region (I). The new BAE in this region are: 
\begin{equation}
\begin{aligned}
 e_1^{2L}(\lambda_j) =& \prod_{k\neq j}^{M_s -1}
        e_2(\lambda_j-\lambda_k) e_2(\lambda_j+\lambda_k) 
        \prod_{\beta=1}^{M_c -2} e_{-1}(\lambda_j-\vartheta_\beta)
        e_{-1}(\lambda_j+\vartheta_\beta)\,,\\
  &\qquad j=1,\ldots,M_s-1,\\
e_{\alpha+2\tau}(\vartheta_\gamma) =&
         e_{2\tau-2-\alpha}(\vartheta_\gamma)
        \prod_{k=1}^{M_s -1}e_{-1}(\vartheta_\gamma-\lambda_k)
                            e_{-1}(\vartheta_\gamma+\lambda_k)\,,
  \quad \gamma=1,\ldots,M_c-2.
\label{bae3}
\end{aligned}
\end{equation}
The bound states in this sector, consisting of two charge and one spin
rapidity can be interpreted as a singlet with vanishing charge bound to the
impurity. 
\end{itemize}

Increasing $\tau$ beyond $\alpha/2 +1$ does not lead to additional bound 
states as expected for an impurity with a single orbital allowing for 
occupation of at most two charges. Therefore, we conclude that the 
maximum number of bound states allowed to develop themselves upon 
variation of the hybridization at the boundary impurity site is three 
- two holons and one spinon.

\subsection{Continuum limit - Equations for the densities}

The analysis of the BAE is simplified by doubling of the real roots of 
the BAE with positive and negative ones identified, i.e. 
$\lambda_{-j} = -\lambda_j$ and $\vartheta_{-\gamma} = 
-\vartheta_\gamma$. In the thermodynamic limit, the real roots 
$\{ \lambda_j \}$ ($\{ \vartheta_\gamma \}$) form continuous 
distributions which are conveniently described in terms of their 
densities $\rho_s (\lambda)$ ($\rho_c (\vartheta)$). Those densities 
obey the following coupled, but linear, integral equations:

\begin{equation}
  \left( \begin{array}{c} \rho_s (\lambda)\\ 
    \rho_c (\vartheta)\end{array} \right) =
  \left( \begin{array}{c} 2 a_1(\lambda)\\ 
    0 \end{array} \right) +
       {1 \over L}\left( \begin{array}{c} \hat{\rho}_s^{(R)}(\lambda) + 
         \hat{\rho}_s^{(b)}(\lambda)\\
         \hat{\rho}_c^{(R)}(\vartheta) + 
         \hat{\rho}_c^{(b)}(\vartheta)\end{array} \right)+
       \left( \begin{array}{cc} -\int_{-A}^A a_2 & \int_{-B}^B a_1 \\ 
         \int_{-A}^A a_1 & 0\end{array} \right)*
         \left( \begin{array}{c} \rho_s (\lambda) \\ 
           \rho_c (\vartheta)\end{array} \right).\ 
         \label{eqdens}
\end{equation}
Here we have introduced $a_y(x)={1 \over 2\pi}\frac{y}{y^2/4+x^2}$ and
$\int_{-k}^k f*g$ denotes the convolution $\int_{-k}^{k}dy f(x-y)g(y)$.
The boundaries of integration $A$ and $B$ for the spin and charge sector,
respectively, are fixed by the conditions
\begin{equation}
\begin{aligned}
   \int_{-A}^{A}\mathrm{d}\lambda\ \rho_s(\lambda) =&
        \frac{2\left[M_s-\theta(\tau-\tau_1)\right]+1}{L}\,,
\\
        \int_{-B}^{B}\mathrm{d}\vartheta\ \rho_c(\vartheta) =&
        \frac{2\left[M_c-\theta(\tau-\tau_0)-\theta(\tau-\tau_2)\right]+1}{L},
        \label{eqfix}
\end{aligned}
\end{equation}
where $\theta(x)$ is the Heaviside step function and $\tau_k$ are the
thresholds for the appearance of bound states identified above. Note that the
boundaries of integration $A$ and $B$ are completely fixed through
(\ref{eqfix}) by bulk quantities (i.e. total densities of spin $\sigma$
electrons).  Alternatively, they can be specified in a grand canonical
approach by the conjugate potentials, i.e. the chemical potential $\mu$ and
the magnetic field $H$.  In this case the ground state is obtained by filling
all modes with negative dressed energies $\varepsilon_s(\lambda)$ and
$\varepsilon_c(\vartheta)$ solving the integral equations
\begin{equation}
\left( \begin{array}{c} \varepsilon_s (\lambda)\\ 
    \varepsilon_c (\vartheta)\end{array} \right) =
  \left( \begin{array}{c} -2\pi a_1(\lambda) + H\\ 
    \mu - {H \over 2} \end{array} \right) +
       \left( \begin{array}{cc} -\int_{-A}^A a_2 & \int_{-B}^B a_1 \\ 
         \int_{-A}^A a_1 & 0\end{array} \right)*
         \left( \begin{array}{c} \varepsilon_s (\lambda) \\ 
           \varepsilon_c (\vartheta)\end{array} \right).\ 
         \label{de}
\end{equation}
with $\varepsilon_s(\pm A)=0$ and $\varepsilon_c(\pm B)=0$.
The unpolarized ground state for vanishing magnetic field, for example,
corresponds to $A=\infty$. 

Since (\ref{eqdens}) is a linear system of integral equations, the generic
solution for both spin and charge densities is of the form 
\begin{equation}
  \rho = \rho_{\infty} + {1 \over L} (\rho_{imp} + \rho_{b}).
  \label{frho}
\end{equation}
The first term in (\ref{frho}) is the bulk density obtained by solving
(\ref{eqdens}) with $L=\infty$.  The remaining two terms of order $1/L$ are
the contributions due to the presence of the impurity and due to the openness
of the chain, respectively.

Depending on the region fixed by the parameters $\alpha$ and $t$ (or $\tau$)
the driving terms $\hat\rho^{(R)}_{s,c}$ in the integral equations
(\ref{eqdens}) are obtained from the discrete BAE (\ref{bae0}), (\ref{bae1}),
(\ref{bae2}) or (\ref{bae3}) as
\begin{equation}
  \hat{\rho}_s^{(R)}(\lambda)= 
  \left\{
  \begin{array}{ll}
    0 & \quad R=0 \\
    a_{2\tau +1 -\alpha}(\lambda)+a_{1+\alpha -2\tau}(\lambda) & 
    \quad R=I\\
    -a_{3+\alpha-2\tau}(\lambda)-a_{2\tau-\alpha-1}(\lambda)& 
    \quad R=II\\
    0& \quad R=III  
  \end{array}\right.
  \label{dtrhos}
\end{equation}
for the spin--sector and
\begin{equation}
  \hat{\rho}_c^{(R)}(\vartheta)= 
  \left\{
\begin{array}{ll}
a_{\alpha}(\vartheta+t) + a_{\alpha}(\vartheta-t) & R=0\quad
                                                 (t=i\tau\in\mathbb{R}) \\
a_{\alpha + 2\tau}(\vartheta) + a_{\alpha-2\tau}(\vartheta) & R=0\quad
                                                 (\tau=-it\in\mathbb{R}) \\
a_{\alpha + 2\tau}(\vartheta) + a_{\alpha-2\tau}(\vartheta) & R=I \\
a_{\alpha + 2\tau}(\vartheta) + a_{2 +\alpha-2\tau}(\vartheta)& R=II,III
\end{array}\right.
\label{dtrhoc}
\end{equation}
for the charge--sector.
The contributions due to the boundaries can be calculated with the driving
terms $\hat{\rho}_{s}^{(b)}(\lambda)=a_2 (\lambda)$ and
$\hat{\rho}_{c}^{(b)}(\vartheta)=-a_1 (\vartheta)$.  As a consequence of the
decomposition (\ref{frho}), the bulk, impurity and boundary contributions to
any thermodynamic quantity can be studied separately.  For instance, the
ground state energy per site is formally
\begin{equation}
  \frac{E_0}{L}=\epsilon_{\infty} + \frac{\epsilon_{imp}+\epsilon_{b}}{L}.
  \label{E0}
\end{equation}
From this, magnetization, density of electrons and the corresponding
susceptibilities are obtained by taking the appropriate derivatives with
respect to the conjugate fields.  The focus of this paper is on the
characterization of the impurity.  Therefore, we will concentrate on the
contributions to these quantities which are determined by (\ref{dtrhos}) and
(\ref{dtrhoc}).  The finite size contributions of the boundaries are of purely
geometric nature and have been studied by E{\ss}ler\cite{Essl96}.

\subsection{The impurity's ground state}

So far, we have distinguished four regions of the $\alpha$--$\tau$ parameter
space of the impurity, characterized by different possible bound state
configurations. Each of theses regions is described by a different set of
BAE. Before studying how the properties of the system are affected by the
presence of the impurity, the configuration corresponding to the true ground
state has to be identified.  For this, the impurity's contribution to the
energy has to be computed using the applicable sets of BAE in the vicinity of
the transition lines, $\tau=\tau_k$.  The formal expression for this
contribution in terms of the densities is
\begin{equation}
\begin{aligned}
  \epsilon_{imp} =& (\alpha +2) V_0 -2
   -\pi \int_{-A}^{A}\mathrm{d}\lambda\ \rho_s^{(R)}(\lambda)a_1(\lambda) + 
\\&
    \left(\mu - \frac{H}{2}\right)\bigg(
    \frac{1}{2}\int_{-B}^{B}\mathrm{d}\vartheta\ \rho_c^{(R)}(\vartheta) + 
    \theta(\tau - \tau_0)+\theta(\tau-\tau_2)\bigg)
  \\&
 +H\left(\frac{1}{2}\int_{-A}^{A}\mathrm{d}\lambda\ \rho_s^{(R)}(\lambda) +
\theta(\tau - \tau_1)\right) -2\mu.
\label{eimp}
\end{aligned}
\end{equation}
Here, we want to concentrate on the case $H=0$: for small magnetic fields
(i.e. $A\gg1$), a convenient treatment of the integral equations
(\ref{eqdens}) is obtained by rewriting the one for $\rho_s$ as follows
\begin{equation}
  \rho_s^{(R)} = \frac{\hat{\rho_s}^{(R)}}{1+a_2} + 
  \left(\int_{-\infty}^{-A} + 
  \int_A^{\infty}\right) \frac{a_2}{1+a_2}*\rho_s^{(R)} + 
  \int_{-B}^{B} \frac{a_1}{1+a_2}*\rho_c^{(R)}.
  \label{Frhos}
\end{equation}
Now, at $H=0$ ($A \to \infty$), the densities are given in terms of the 
solution to a scalar integral equation for the density $\rho_c^{(R)}$ 
of charge rapidities:
\begin{equation}
  \begin{aligned}
  \rho_s^{(R)} =&  \hat{\rho}_{s,H=0}^{(R)} + \int_{-B}^B G_0*\rho_c^{(R)}\\ 
  \rho_c^{(R)} =&   \hat{\rho}_{c,H=0}^{(R)} + \int_{-B}^B G_1*\rho_c^{(R)}\ .
  \end{aligned}
 \label{eqdensH=0}
\end{equation}
The driving terms at zero field are
\begin{equation}
  \hat{\rho}_{s,H=0}^{(R)}(\lambda)= 
  \left\{
  \begin{array}{ll}
    0 & \quad R=0 \\
    G_{2\tau -\alpha}(\lambda) + G_{\alpha -2\tau}(\lambda) & \quad R=I\\
    -G_{2+\alpha-2\tau}(\lambda) - G_{2\tau-\alpha -2} (\lambda)& 
    \quad R=II\\
    0& \quad R=III  
  \end{array}\right.
  \label{dtrhosH0}
\end{equation}
and
\begin{equation}
  \hat{\rho}_{c,H=0}^{(R)}(\vartheta)= 
  \left\{
  \begin{array}{ll}
    \hat{\rho}_c^{(0)} & \quad R=0 \\
    \hat{\rho}_c^{(I)}+G_{2\tau -\alpha +1}(\vartheta)+
    G_{\alpha -2\tau +1}(\vartheta))  & \quad R=I\\
    \hat{\rho}_c^{(II)} -G_{3+\alpha-2\tau}(\vartheta)-
    G_{2\tau-\alpha-1}(\vartheta)& \quad R=II\\
    \hat{\rho}_c^{(III)}& \quad R=III  
  \end{array}\right.
  \label{dtrhocH0}
\end{equation}
for all four regions. The functions $\hat{\rho}_c$ have been defined above 
in (\ref{dtrhoc}) and
\begin{equation}
  G_{\beta}(x) = \int_{-\infty}^{\infty}{\mathrm{d}\omega \over 2\pi}
  \exp(-i\omega x)\frac{\exp(-\beta|\omega/2|)}{2\cosh(\omega/2)}.
  \label{Gfunc}
\end{equation} 
Now, we can express the impurity's contribution (\ref{eimp}) to the energy 
in terms of the solution to the integral equation (\ref{eqdensH=0}).
\begin{itemize}
\item[{\bf (0)}:]{$t=i\tau$ real and $0<\tau<\tau_0$}

 In the absence of bound state, the $1/L$ correction to the ground state 
 energy reads
  \begin{equation}
    \epsilon_{imp}^{(0)} = E_b + \frac{1}{2}\int_{-B}^{B}\mathrm{d}\vartheta\  
            [\mu - 2\pi G_1 (\vartheta)]\rho_c^{(0)}(\vartheta) ,
    \label{eimp0}
  \end{equation}
where $E_b = V_0 (\alpha+2)- 2(\mu +1)$.  \remark{The term $E_b$ is given by
  (\ref{Eb}) and is the energy of the geometric boundaries. It does play a
  role in the discussion of the various phases of the system since it is
  ($\alpha$,$\tau$) dependent.}

\item[{\bf (I)}:]{$\tau_0<\tau<\tau_1$}

  As we have seen previously, in region (I) a first bound state is created 
  in the charge sector. The energy contribution due to the impurity now 
  becomes:
  \begin{equation}
    \epsilon_{imp}^{(I)} = E_b -\pi [G_{2\tau -\alpha +1}(0)+
      G_{\alpha -2\tau +1}(0)]
    + \frac{1}{2}\int_{-B}^{B} \mathrm{d}\vartheta\ 
    [\mu - 2\pi G_1 (\vartheta)] \rho_c^{(I)}(\vartheta) +\mu .
    \label{eimp1}
\end{equation}
  Here $\rho_c$ has to be evaluated with the appropriate driving term 
  (\ref{dtrhocH0}) for region (I). 
  The additional chemical potential is due to the charge in the bound state
  and the terms containing the $G$-functions are the consequence of the 
  rearrangement of the rapidities in the spin sector.  
\item[{\bf (II)}:]{$\tau_1<\tau<\tau_2$}

  In region (II), two bound states are possible. Using the appropriate 
  driving terms in (\ref{eqdensH=0}), the $1/L$ correction to the energy 
  becomes:
  \begin{equation}
    \epsilon_{imp}^{(II)} = E_b +\pi [G_{3+\alpha -2\tau}(0)+
      G_{2\tau -\alpha-1}(0)]
    + \frac{1}{2}\int_{-B}^{B} \mathrm{d}\vartheta\ 
    [\mu - 2\pi G_1 (\vartheta)]\rho_c^{(II)}(\vartheta) +\mu+E_{M_s}.
    \label{eimp2}
  \end{equation}
  The extra term, $E_{M_s}$, is the energy contribution (\ref{ebound_s}) 
  of the spin bound state.

\item[{\bf (III)}:]{$\tau>\tau_2$}
  
  In region (III), $\epsilon_{imp}$ takes the form
  \begin{equation}
    \epsilon_{imp}^{(III)} 
        = E_b + \frac{1}{2}\int_{-B}^{B}\mathrm{d}\vartheta\ 
 [\mu - 2\pi G_1 (\vartheta)]\rho_c^{(III)}(\vartheta) +
    2\mu+E_{M_s}.
    \label{eimp3}
  \end{equation}
\end{itemize}

\begin{figure}[ht]
\includegraphics[width=0.7\textwidth]{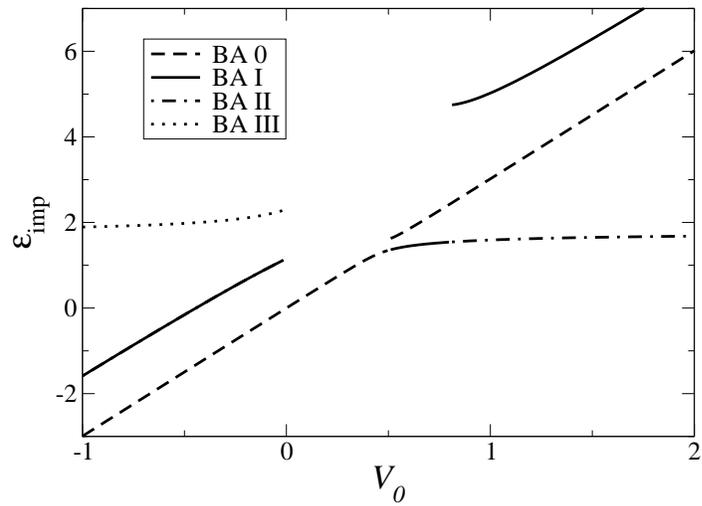}
\caption{
\label{Etau}
Impurity contribution to the ground state energy for different bound state
configurations as a function of the hybridization $V_0$ for bulk hole
concentration of $\delta=0.2$.  The impurity parameter is fixed to
$\alpha=1$.  Note that the Bethe ansatz for the configuration $R=I$ and $R=II$
gives identical results for $\varepsilon_{imp}$ at negative $V_0$.}
\end{figure}

The proper ground state configuration is the one which minimizes the impurity
contribution to the energy as given in (\ref{eimp0})--(\ref{eimp3}).  In
Fig. \ref{Etau}, we present numerical data for $\epsilon_{imp}$, for fixed
$\alpha=1$ and bulk hole concentration $\delta=0.2$, as a function of the
hybridization $V_0$ (note that $\epsilon_{imp}^{(R)}$ can be continued to
regions with larger $R$, describing a configuration where an allowed bound
state is not occupied).
From these numerical data we conclude that -- as far as the ground state of
the impurity is concerned -- three intervals in the hybridization have to be
distinguished (see also Fig.~\ref{V0aplane}):
\begin{description}
\item $V_0<1/(\alpha+1)$ (real $t$ and $\tau=-it \in [0,\tau_0] \cup
  [\tau_2,\infty]$): for attractive ($V_0<0$) and weakly repulsive ($V_0>0$)
  hybridization the bound states identified above are not occupied, hence the
  ground state is described by Eqs.~(\ref{eqdens}) with $R=0$.

\item $1/(\alpha+1)<V_0<4/(2\alpha+3)$ (i.e.\ $\tau_0<\tau<\tau_1$):  in this
  interval the holon bound state is occupied in the ground state configuration
  is described by (\ref{eqdens}) with the driving terms for $R=I$.

\item $V_0>4/(2\alpha+3)$ (i.e.\ $\tau_1<\tau<\tau_2$): for strong repulsive
  coupling between the impurity and the host the ground state is obtained
  using the configuration from $R=II$, i.e.\ with the holon \emph{and} the
  spinon bound state occupied.
\end{description}
Note that the charge $Q=0$ singlet bound state of region III is never present
in the ground state of the system.

In the following we continue to label our results by the index $R=0$ to $III$,
the relation to the physical ground state as a function of $V_0$, however, is
given by the classification above.

\section{Zero-field susceptibility and occupation of the impurity}%
\label{sec:ZeroH}

\subsection{Analytical results close to half-filling}

In this section,we will explicitly calculate the magnetization of the impurity
in a small field.  With our parameterization of the BAE's roots, the impurity
contribution to the magnetization is given by
\begin{equation}
  M_{imp} = 
  \frac{1}{4}\int_{-B}^{B}\mathrm{d}\vartheta\  \rho_c^{(R)} (\vartheta) 
       +{1 \over 2}(\theta(\tau-\tau_0)
       +\theta(\tau-\tau_2))-
       \frac{1}{2}\int_{-A}^{A}\mathrm{d}\lambda\ \rho_s^{(R)} (\lambda)-
       \theta(\tau-\tau_1).
       \label{Mimp}
\end{equation}
Proceeding as for Eq.~(\ref{Frhos}), we can rewrite this expression for the
impurity's magnetization as an integral over the spin density only:
\begin{equation}
  M_{imp}={1 \over 2}\int_{A}^{\infty}\mathrm{d}\lambda\ \rho_s^{(R)} (\lambda)
  -\theta(\tau-\tau_1)
  + {1 \over 2}(\theta(\tau-\tau_0)
  +\theta(\tau-\tau_2)).
  \label{Mimp2}
\end{equation}
Introducing $g(z)=\rho_s^{(R)}(A+z)$ in the integral equation (\ref{Frhos}),
we obtain the following equation for the unknown function $g$:
\begin{equation}
\begin{aligned}
  g(z)&= \hat\rho_{s,H=0}^{(R)}(A+z) + 
  \int_0^{\infty}\mathrm{d}z'\ G_1 (z-z')g(z')\\
  &+\int_0^{\infty}\mathrm{d}z'\  G_1 (2A+z+z')g(z')
  +\int_{-B}^{B}\mathrm{d}z'\  G_0 (z-z' +A) \rho_c^{(R)} (z').
  \label{geq}
  \end{aligned}
\end{equation}
For small magnetic fields corresponding to large values of $A$ this equation
can be solved by iteration using Wiener-Hopf methods.  Following
Ref.~\onlinecite{YaYa66}, we expand $g=g_1 + g_2 + ...$, where
\begin{eqnarray}
  g_1 (z) &=& g_0^{(R)}(A+z) + 
  C_{\alpha,\tau}^{(R)} G_0 (A+z)
  + \int_0^{\infty}\mathrm{d}z'\  G_1(z-z')g_1(z')
  \label{g1}\\
  g_n (z) &=& 
  \int_0^{\infty}\mathrm{d}z'\  G_1 (2A +z+z') g_{n-1} (z') +
  \int_0^{\infty}\mathrm{d}z'\  G_1(z-z')g_n (z')\ , \quad n >1.
\label{gn}
\end{eqnarray}
Here $g_0^{(R)} =\hat{\rho}_{s,H=0}^{(R)}$ and we have used that, for $A \gg
B$,
\begin{equation}
\int_{-B}^{B}\mathrm{d}z'\  G_0 (A+z-z')  \rho_c^{(R)} (z')\approx 
G_0 (A+z)\int_{-B}^{B}\mathrm{d}z'\  e^{\pi z'}\rho_c^{(R)}(z')
\end{equation}
to define the number $C_{\alpha,\tau}^{(R)}\equiv \int_{-B}^B\mathrm{d}z'\
e^{\pi z'}\rho_c^{(R)}(z')$ which is given in terms of $\rho_c^{(R)}$ alone.

The leading behaviour of the impurity magnetization for large $A$ is now
obtained from (\ref{g1}): using the results (\ref{WH_ga}) and (\ref{gbpS})
from Appendix~\ref{appWH} we find
\begin{equation}
  M_{imp} =
  \frac{e^{-\pi A}}{\sqrt{2\pi e}}
 \left\{
 \begin{array}{ll}
   C_{\alpha,\tau}^{(0)}  \\
   C_{\alpha,\tau}^{(I)} + 2\cos({\pi(2\tau-\alpha) \over 2}) \\
   C_{\alpha,\tau}^{(II)} + 2\cos({\pi(2\tau-\alpha) \over 2})\\
 \end{array}\right\}
 \begin{array}{ll}
   & \quad R=0\\
   +1/2 & \quad R=I\\
   -1/2 & \quad R=II\\
 \end{array}
 \label{MAimp}
\end{equation}

Here the dependence of $M_{imp}$ on the density of electrons in the host is
completely given through the constants $C_{\alpha,\tau}$.  The zero field
limit of these quantities is given in terms of the solution to the integral
equation (\ref{eqdensH=0}) for $\rho_c^{(R)}$.  In general this equation has
to be solved numerically.
For $B \ll 1,|2\tau-\alpha|$ (i.e.\ close to half-filling and the impurity
sufficiently far away from the threshold for the holon bound state), however,
it can be solved by iteration.  Doing so, we obtain at first order in $B$ the
following expressions for $C_{\alpha,\tau}$:
\begin{equation}
  C_{\alpha,\tau}^{(R)}= 2B
  \left\{
  \begin{array}{ll}
    \frac{4\alpha}{\pi(\alpha^2-4\tau^2)}& \quad R=0 \\
    \frac{4\alpha}{\pi(\alpha^2-4\tau^2)} +G_{2\tau+1-\alpha}(0)
    +G_{1+\alpha-2\tau}(0)& \quad R=I\\
    \frac{4(1+\alpha)}{\pi(2+\alpha-2\tau)(\alpha+2\tau)} 
    -G_{3+\alpha-2\tau}(0)
    -G_{2\tau-\alpha-1}(0)& \quad R=II\\
  \end{array}\right.
  \label{CR}
\end{equation}
At the threshold, $\tau=\tau_0=\alpha/2$, the leading contribution to
$C_{\alpha,\tau_0}^{(0)}$, $C_{\alpha,\tau_0}^{(I)}$ at small hole
concentration are $\pm 1$ independent of $\delta$.

Finally, the boundaries of integration, $A$ and $B$, have to be expressed in
terms of physical quantities, namely the concentration $\delta=M_c/L$ of holes
(doping) in the bulk and the magnetic field using the relations (\ref{eqfix})
and (\ref{de}). Again, restricting ourselves to the regime close to half
filling we have $\pi\delta = 2B \ln 2 $.  To express $A$ in terms of the
magnetic field one has to enforce $\varepsilon_s(A)=0$.  A Wiener-Hopf
analysis of the integral equations (\ref{de}) gives\cite{Essl96}
\begin{equation}
  \pi A= -\ln(H/H_0)+\frac{1}{4\ln H}, \quad H_0 = \sqrt{(2\pi/e)}(2\pi-C)
  \label{AH}
\end{equation}
with $C = \int_{-B}^B \mathrm{d}\vartheta\ e^{\pi\vartheta} \varepsilon_c
(\vartheta)$, here $\varepsilon_c$ is the dressed energy of the holes given by
(\ref{de}).  Close to half-filling $C \ll 2\pi$ and $H_0\approx
(2\pi)^{3/2}/\sqrt{e}$.
Using (\ref{CR}) and (\ref{AH}) the low-field magnetization is found to be
linear in $H$.  Thus we obtain for the impurity contribution to the zero field
magnetic susceptibility close to half filling close to half filling $\delta
\ll 1,|2\tau-\alpha|$:
\begin{itemize}
\item[{\bf (0)}:]
  \begin{equation}
    \chi_{imp} = \frac{1}{\pi^2}\frac{\alpha \delta}{(\alpha^2-4\tau^2)
      \ln 2}+ {\cal O}(\delta^2)
    \label{chi0}
  \end{equation}
\item[{\bf (I)}:]
  \begin{equation}
    \begin{aligned}
      \chi_{imp} =& \frac{1}{2\pi^2}\Bigg[\frac{\pi\delta}{2\ln 2}
        \bigg(\frac{4\alpha}{\pi(\alpha^2-4\tau^2)} +G_{2\tau+1-\alpha}(0)
        +G_{1+\alpha-2\tau}(0)\bigg)\\
        &+\cos\left({\pi(2\tau-\alpha) \over 2}\right)\Bigg]
      + {\cal O}(\delta^2)
      \label{chi1}
    \end{aligned}
  \end{equation}
\item[{\bf (II)}:]
  \begin{equation}
    \begin{aligned}
      \chi_{imp} =& \frac{1}{2\pi^2}\Bigg[\frac{\pi\delta}{2\ln 2}
        \bigg(\frac{4(1+\alpha)}{\pi(2+\alpha-2\tau)(\alpha+2\tau)} 
              -G_{3+\alpha-2\tau}(0) -G_{2\tau-\alpha-1}(0)\bigg)\\
        &+\cos\left({\pi(2\tau-\alpha) \over 2}\right)\Bigg]
      + {\cal O}(\delta^2)
      \label{chi2}
    \end{aligned}
  \end{equation}
\end{itemize}

Finally, we can compute the number of electrons on the impurity which is
obtained from (\ref{eqfix}) to be
\begin{equation}
D_{imp} = 
  2 - \frac{1}{2}\int_{-B}^{B}\mathrm{d}\vartheta\ \rho_c^{(R)} (\vartheta)
  -\theta(\tau-\tau_0)-\theta(\tau-\tau_2).
  \label{Dimpdef}
\end{equation}
Again this expression can be calculated at zero field, close to half-filling
and away from the holon bound state threshold ($B \ll 1,|2\tau-\alpha|$) giving
\begin{equation}
  D_{imp}=
  \left\{
  \begin{array}{ll}
    2-\frac{\pi\delta}{\ln 2}\frac{2\alpha}{\pi(\alpha^2-4\tau^2)}&
               \quad R=0 \\
    1- \frac{\pi\delta}{2\ln 2}\left(\frac{4\alpha}{\pi(\alpha^2-4\tau^2)}
               +G_{2\tau+1-\alpha}(0) 
    +G_{1+\alpha-2\tau}(0)\right)& \quad R=I\\
    1-\frac{\pi\delta}{2\ln
               2}\left(\frac{4(1+\alpha)}{\pi(2+\alpha-2\tau)(\alpha+2\tau)}  
    -G_{3+\alpha-2\tau}(0)
    -G_{2\tau-\alpha-1}(0)\right)& \quad R=II\\
  \end{array}\right.
  \label{Dimp}
\end{equation}

\subsection{Numerical results for arbitrary doping}
\begin{figure}[ht]
\includegraphics[width=0.7\textwidth]{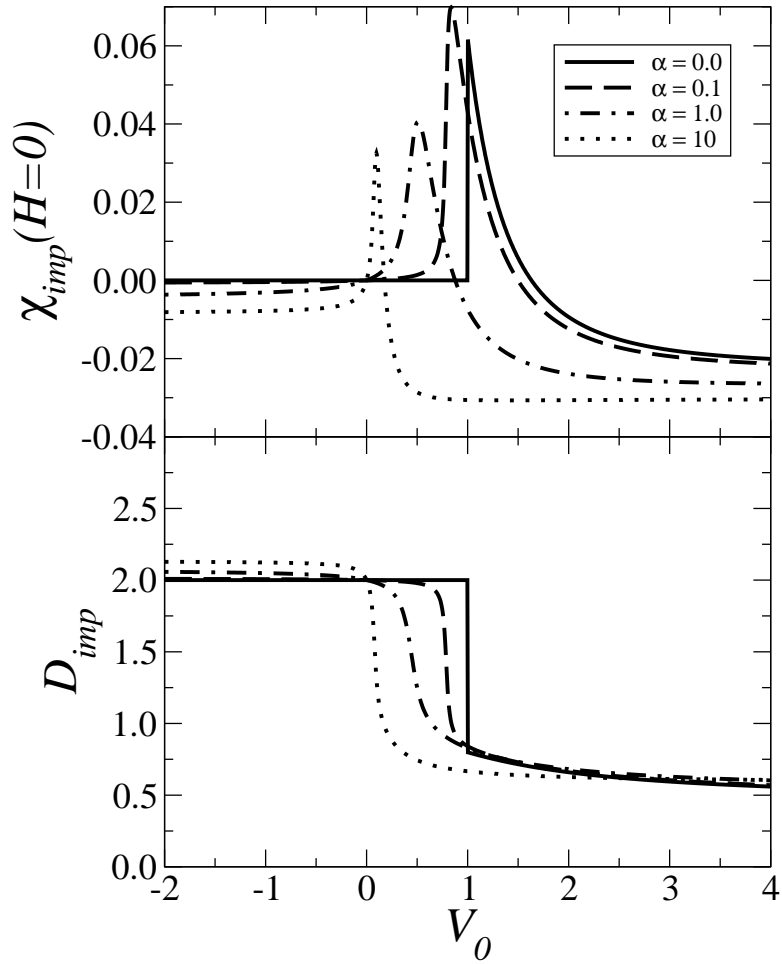}

  \caption{
    \label{chi0_02}
    Zero field susceptibility $\chi_{imp}$ (upper panel) and electron number
    $D_{imp}$ on the impurity site (lower panel) for bulk doping $\delta =
    0.2$ as a function of $V_0$.  }
  \end{figure}

\begin{figure}[ht]
\includegraphics[width=0.7\textwidth]{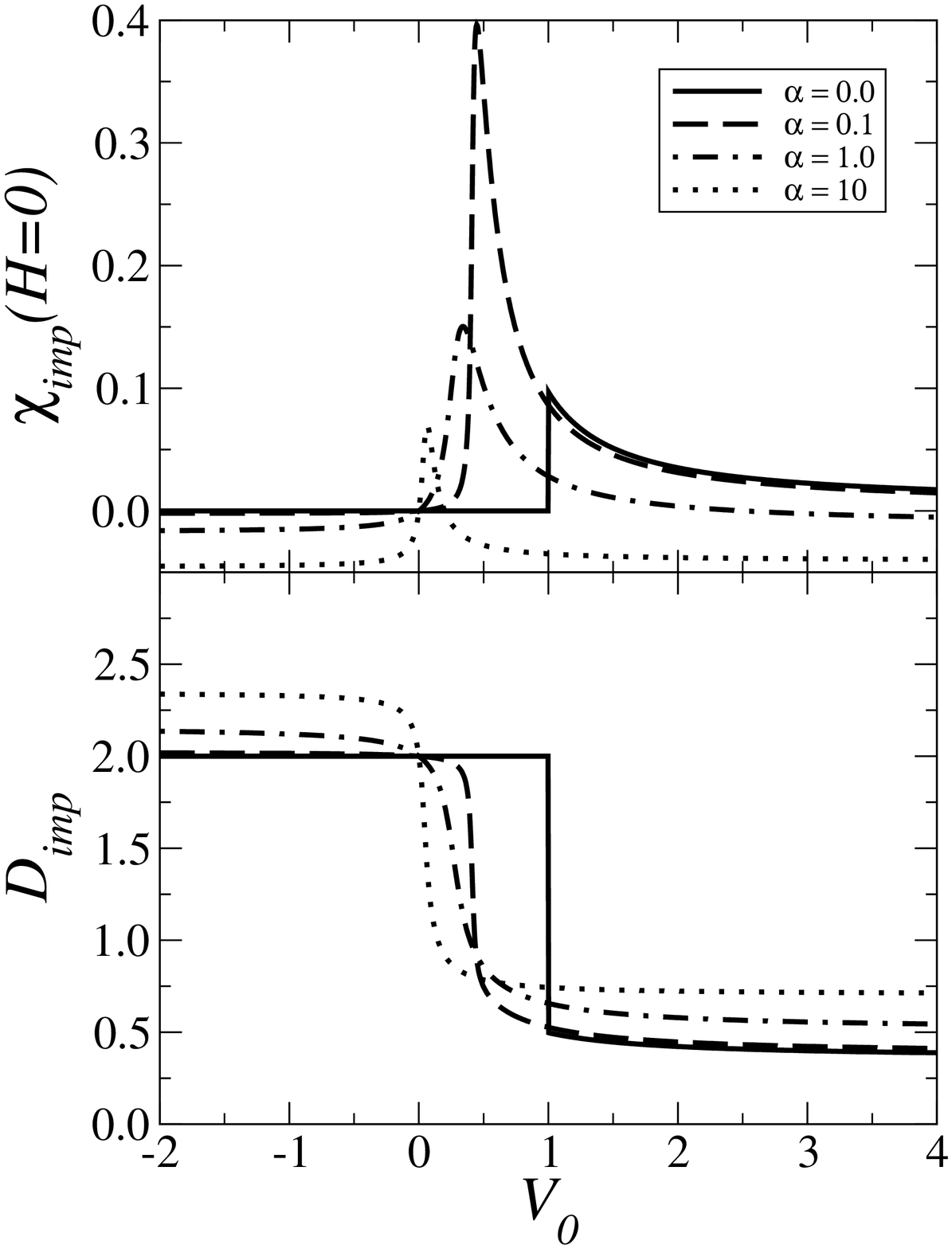}
  \caption{
    \label{chi0_05}
    Same as Fig.~\ref{chi0_02} for $\delta=0.5$.
        }
  \end{figure}

\begin{figure}[ht]
\includegraphics[width=0.7\textwidth]{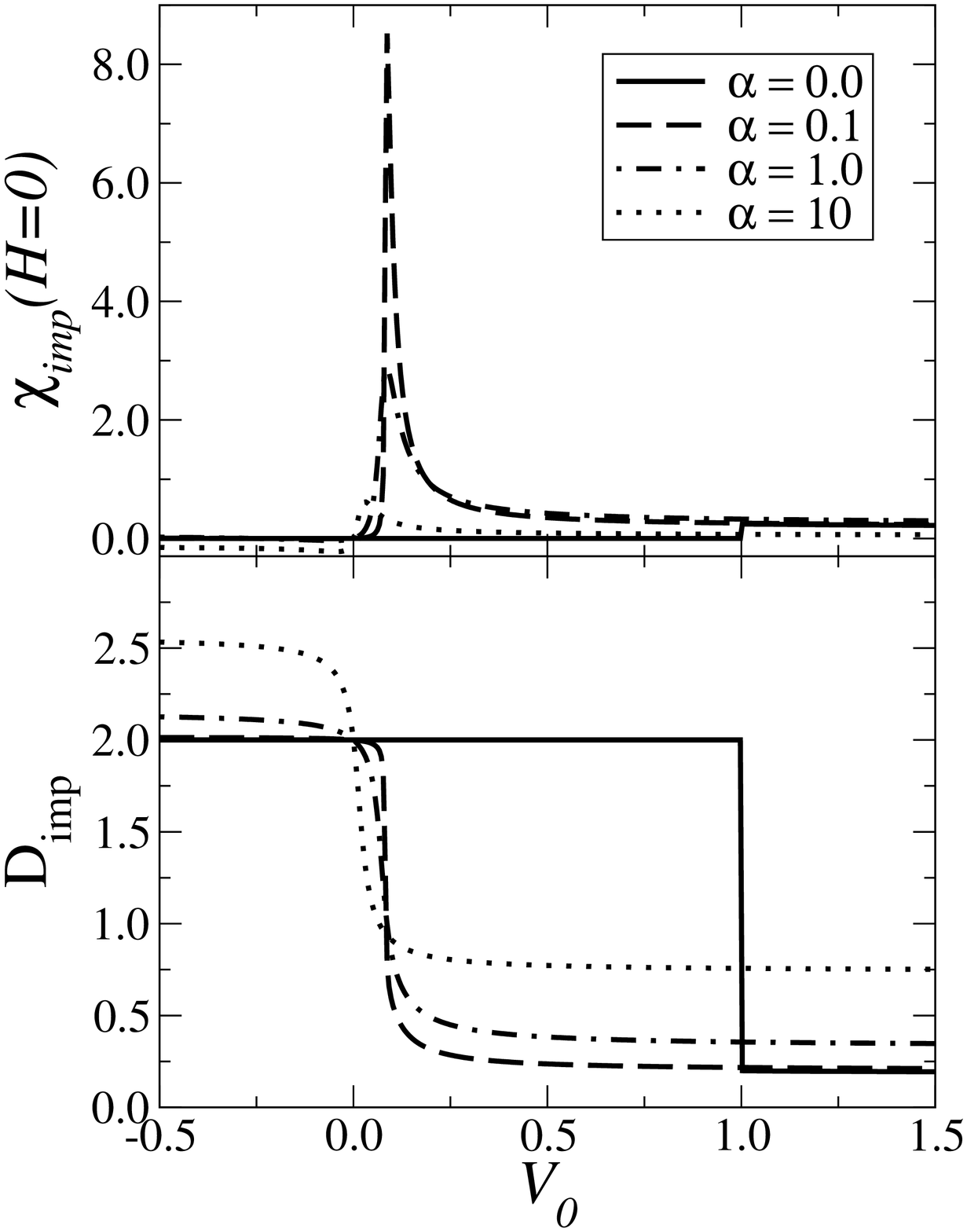}
  \caption{
    \label{chi0_08}
     Same as Fig.~\ref{chi0_02} for $\delta=0.8$.  }
\end{figure}

For finite density of holes ($B$ finite) the integral equations
(\ref{eqdensH=0}) and (\ref{de}) for the charge components of the densities
and dressed energies have to be solved numerically.
In Figures~\ref{chi0_02}, \ref{chi0_05} and \ref{chi0_08}, we present results
of this analysis for the zero field susceptibility and occupation of the
impurity for different doping as a function of the hybridization $V_0$.  Note
that for general filling factors, the constant $C$ in the definition
(\ref{AH}) of $H_0$ is no longer negligible, thus
\begin{equation}
  \chi_{imp}(H=0) =
  \frac{1}{2\pi (2\pi-C)}
 \left\{
 \begin{array}{ll}
   C_{\alpha,\tau}^{(0)}&\quad R=0 \\
   C_{\alpha,\tau}^{(I)} + 2\cos({\pi(2\tau-\alpha) \over 2})&\quad R=I\\
   C_{\alpha,\tau}^{(II)} + 2\cos({\pi(2\tau-\alpha) \over 2})&\quad R=II\\
 \end{array}  \right.
 \label{ChiH=0}
\end{equation}

First notice that the $\alpha=0$ case where the doubly occupied state
decouples from the remaining three electronic impurity states $|0\rangle$,
$|\sigma\rangle$ is special: here, for $V_0 <1$, the impurity is doubly
occupied $D_{imp}\equiv2$.  Therefore the impurity does not contribute to the
magnetic susceptibility of the system, i.e.\ $\chi_{imp}=0$.  For $V_0 >1$
(regions (I) and (II)), the occupation is less than $2$ due to the filled
holon bound state and there is a small finite impurity contribution to the
susceptibility.

For $\alpha > 0$, the total number of electrons present at the impurity can
become larger than $2$ as a consequence of the attractive interaction between
the impurity and the bulk electrons for $V_0<0$.  Note, however, that the
difference $D_{imp}|_{V_0=-\infty}-D_{imp}|_{V_0=+\infty}$ approaches $2$,
showing the depletion of the impurity orbital at large positive $V_0$.
For small hole concentration $\delta$ we had found above that the impurity
occupation changes at the threshold for the formation of the holon bound state
$V_0\agt 1/(\alpha+1)$.  For finite $\delta$ this resonance moves to smaller
values of $V_0$, determined by the condition that the (real) impurity
parameter $t$ is of the order of the boundary $B$ which is given by the bulk
hole concentration through Eq.~(\ref{eqdens}).  Below this value of the
hybridization, i.e.\ for $t\agt B$ the impurity is doubly occupied and
essentially decoupled from the host.

\begin{figure}[ht]
\includegraphics[width=0.7\textwidth]{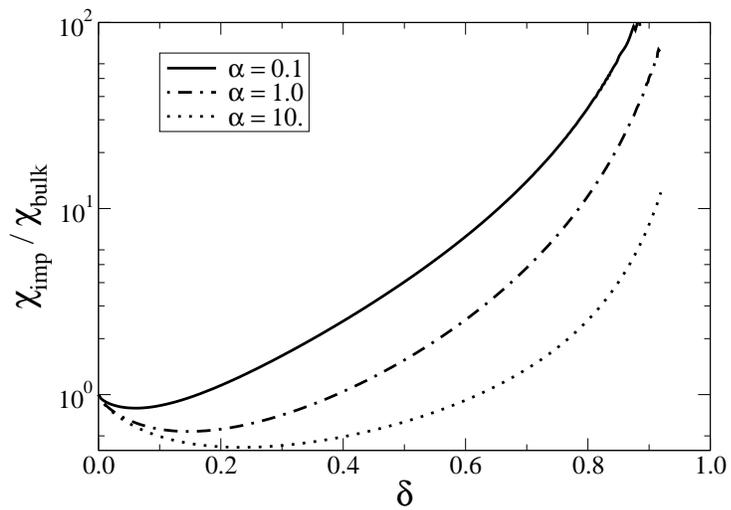}
  \caption{
    \label{fig:maxchi0}
    Maximum value of $\chi_{imp}(H=0)$ as a function of $\delta$.  For small
    $\delta$ this maximum appears close to the threshold for the creation of
    the holon bound state $V_0=1/(\alpha+1)$.}
  \end{figure}

The same shift is observed in the resonance of the impurity contribution to
the zero-field susceptibility which moves from the threshold for the holon
bound state towards smaller values of $V_0$ as the hole concentration is
increased.
This resonance is the response of the unpaired electron on the impurity site
which appears when $D_{imp}\approx1$.  Although still limited by fluctuations
in the impurity's occupation the susceptibility at the resonance grows
strongly with $\delta$.  This is shown in Figure~\ref{fig:maxchi0} where the
maximum of $\chi_{imp}(H=0)$ as a function of $V_0$ is given for different
values of the impurity parameter $\alpha$ as a function of the hole
concentration in the bulk $t$--$J$ chain.  For $\delta\to0$ the susceptibility
approaches that of the bulk system while it diverges for $\delta\to1$, i.e.\
vanishing bulk density of electrons.  In this limit the remaining electron on
the impurity site is essentially an uncoupled local moment.

\section{Impurity in a magnetic field}
%
As we have seen above the magnetic behaviour of the impurity is most
interesting in the regime of weak coupling $V_0$ parameterized by real values
of $t$.  This parameter introduces a scale where the response of the impurity
to external fields is expected to change.  In the present problem where the
impurity site is coupled to the $t$--$J$ chain with separated spin and charge
degrees of freedom this response will depend on the relation of this scale to
the typical energies in these sectors as well.  First, using the relation
(\ref{AH}), we obtain the Kondo field determining the scale where the
susceptibility of an impurity coupled to the spin degrees of freedom changes
as
\begin{equation}
  H_K \sim H_0 \exp(-\pi t) 
      \approx H_0 \exp\left(-\frac{\pi}{\sqrt{V_0}}\right)\,
\label{Hkondo}
\end{equation}
for large $t$ corresponding to $V_0\ll1$.  Note that the dependence of $H_K$
on the hybridization is exponential and not a power-law as found for a Kondo
impurity in a TLL\cite{LeTo92,FuNa94}.  This reflects the absence of
backscattering in the integrable impurity model considered here.

On the other hand, as seen from the BAE (\ref{bae0}), the primary effect of
the impurity (\ref{hamb}) is to induce a phase shift in the charge sector.
Hence, the hybridization has to exceed a minimal value below which the
impurity is doubly occupied and will not contribute to the magnetic response
of the system (this is different from the Anderson impurity model considered
by Bortz \emph{et al.} where the local orbital can be populated by at most one
electron\cite{BoKS05}).  Finally, for the impurity to be visible in the
\emph{magnetic} response, spin and charge sectors have to be coupled at the
scale introduced by the hybridization.  This coupling is determined by the
relative size of the impurity parameter $t$ and that of the host parameters
$A$, $B$.  Note that the latter are functions of the magnetic field: we have
already used that $A\to\infty$ while $B$ approaches a finite constant for
$H\to0$.  On the other hand, the system becomes completely polarized for
$H=H^{sat}=4\cos^2 (\pi\delta/2)$.  In this limit $A\to A^{sat}=
\frac{1}{2}\sqrt{4/H^{sat}-1}= \frac{1}{2} \tan({\pi\delta}/2)$ while
$B\to\infty$.  Therefore, depending on the relative size of $A$ and $B$ two
scenarios can be distinguished for fixed hole concentration $\delta$ in the
bulk:

(i) For sufficiently weak hybridization, corresponding to large $t$ such that
$t=B>A$ at some value $H=H_B$ of the magnetic field, the impurity is decoupled
from the (charge degrees of the) host at the Kondo scale (\ref{Hkondo}).  In
this case the impurity contribution $\chi_{imp}$ to the susceptibility will
exhibit a resonance at $H=H_B>H_K$ while being suppressed below $H_B$.  

(ii) If the impurity and the host are already coupled at the Kondo scale,
i.e.\ $t=A<B$, Kondo like behaviour of the susceptibility can be observed
where the susceptibility approaches some non-universal finite value as
$H\to0$.

Finally, for $V_0$ large enough for $H_K>H^{sat}$ corresponding to $t<A^{sat}$
(or $t$ purely imaginary) no resonance is present in the susceptibility due to
the finite bandwidth of the lattice model.

\begin{figure}[ht]
\includegraphics[width=0.7\textwidth]{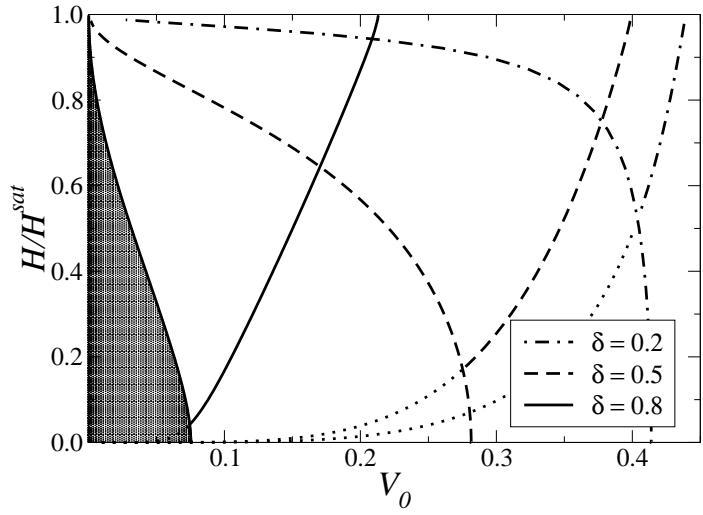}
\caption{
  \label{fig:Hscales}
Relevant energy scales for the impurity problem as a function of the
hybridization $V_0$ for $\alpha=1$: for fixed hole condentration $\delta$
(e.g.\ full lines for $\delta=0.8$) the impurity is decoupled from the host
for magnetic fields below the left line (shaded area).  Above this threshold
(or at sufficiently large $V_0$) the Kondo scale $H_K$ (right branch of full
line) becomes visible in the impurity's response to the external field.  The
dotted lines indicate the continuation of the Kondo scale $H_K$ (\ref{Hkondo})
into the decoupled region.}
\end{figure}

\begin{figure}[ht]
\includegraphics[width=0.7\textwidth]{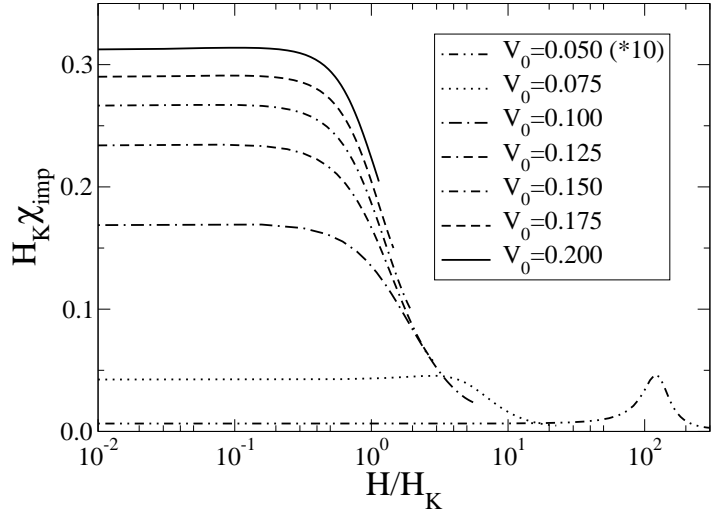}
  \caption{
    \label{fig:chiK}
  Magnetic field dependence of the impurity susceptibility for $\delta=0.8$,
  $\alpha=1$ and various values of the hybridization: for $V_0\agt0.1$ the
  transition between the strong coupling behaviour for small fields and the
  formation of a local moment above $H\approx H_K$ is clearly seen.  At
  smaller values of $V_0$ the susceptibility is strongly suppressed due to the
  decoupling of the doubly occupied impurity from the host (data for
  $V_0=0.05$ are enhanced by a factor of $10$).}
  \end{figure}

In Figure~\ref{fig:Hscales} the regions corresponding to these scenarios are
shown for various values of the hole concentration $\delta$.  By numerical
integration of the Bethe ansatz equations (\ref{eqdens}) the occupation of the
impurity in these regions is determined.  Fluctuations in the densities are
measured by the impurity contribution to the magnetic susceptibility
$\chi_{imp}$ and charge compressibility $\kappa_{imp}=\partial
D_{imp}/\partial \mu$ of the system.  Just as the corresponding bulk
quantities these thermodynamic coefficients are conveniently computed based on
their representation in terms of the dressed charge matrix (see
Appendix~\ref{appDC}).
In Fig.~\ref{fig:IMP} our numerical results on these quantities as a function
of the hybridization and the magnetic field are shown for hole concentration
$\delta=0.8$.  
For intermediate values of the hybridization $0.1\alt V_0\alt 0.2$ the
coupling of the impurity to the holon excitations is effective.  For small
magnetic fields below the Kondo scale $H_K$ the impurity contribution to the
susceptibility takes a non-universal value $\chi_0$ characteristic for the
strong coupling regime of an Anderson impurity.  Above $H_K$ the field
dependence is that of a local moment with logarithmic deviations from full
polarization (see Figure~\ref{fig:chiK}).  
The emergence of universal Kondo like behaviour $\chi_{imp}=f(H/H_K)/2\pi H_K$
for smaller hybridization $V_0\alt0.1$ is suppressed by the decoupling of the
impurity from the host.  Here the double occupancy of the impurity for small
fields and the formation of a local moment with an impurity occupation close
to $1$ appears for fields $H\agt H_B\gg H_K$ are clearly visible.  In the
vicinity of the transition between these regions the occupation and
magnetization of the impurity fluctuate strongly, as shown by the resonances
in the susceptibilities (Fig.~\ref{fig:IMP}).
Finally, for larger values of $V_0\agt 0.2$, the occupation of the impurity
decreases well below $1$ and the susceptibilities are approximately constant
over the entire range of the external field $0<H<H^{sat}$.

\section{Summary \& Conclusion}
We have studied the properties of an Anderson-type impurity embedded into a
supersymmetric $t$--$J$ chain with open boundaries.  Within this model the
hybridization coupling $V_0$ of the impurity to the holons in the 1D host can
be tuned freely while preserving the integrability.  Upon variation of $V_0$
the nature of the many-particle ground state changes due to the appearance of
a sequence of bound states in the holon and spinon sectors.  
From our analysis of the contribution of the impurity to the electronic
density and magnetization of the system and the corresponding susceptibilities
at vanishing magnetic field we have identified two regimes: for attractive or
weakly repulsive hybridization the impurity is doubly occupied with vanishing
fluctuations.  Here the impurity and the host are effectively decoupled due to
a mismatch of $V_0$ and the relevant scales in the holon sector.  For strong
hybridization one holon and one spinon bound state are occupied.  Between
these regions the susceptibility exhibits a sharp resonance which diverges as
the density of bulk electrons tends to $0$.
In a magnetic field exceeding the characteristic scale for excitations in the
spinon sector, $H\agt H_K$, a third region appears for intermediate values of
the hybridization: this regime is characterized by the formation of a local
moment with small corrections to full polarization.  In this range of $V_0$
the field dependence of the susceptibility approaches the characteristic
Kondo scaling behaviour until the decoupling of the impurity
from the holon sector sets in.

Note that the zero field analysis of Section~\ref{sec:ZeroH} can be extended
to small magnetic fields.  This will lead to additional logarithmic
corrections to the thermodynamic quantities.  In the response of the
Hamiltonian (\ref{ham}) they will be hidden by the ones due to the presence of
open boundaries\cite{Essl96}.

The model considered in this paper can be generalized in a number of ways.  As
discussed in the introduction, the boundary impurity can be combined with a
local potential or magnetic field which by itself will generate a sequence of
bound states in the holon and spinon sector.  Fine-tuning the parameters
describing the impurity and those of the boundary field, part of the impurity
spectrum can be projected out from the Hilbert space\cite{FrSl99}.  This
procedure leads, e.g., to exactly solvable models for a true Kondo spin $S$ or
an impurity coupled only to one spin channel of the host electrons (see also
Refs.~\onlinecite{tJKondo,Zhou99}).  While the projection onto a subset of
impurity states is well understood on the level of the Hamiltonian and its
construction it is an open problem, how the corresponding spectra (which will
differ for different choices of the projected subset) are to be computed.  In
particular the role of the bound states in these sectors needs further
investigation which is left for future work.

\begin{acknowledgments}
The authors thank W.~Apel, F.~Essler and M.~Vojta for stimulating discussions and
interest in this work.  This research has been supported by the Deutsche
Forschungsgemeinschaft.
\end{acknowledgments}

\appendix 

\section{Solution of the integral equations (\ref{gn})}
\label{appWH}
The integral equations (\ref{gn}) are of Wiener-Hopf (WH) type
\begin{equation}
  g(z) = g_0(z) + \int_0^{\infty}\mathrm{d}z'\  G_1(z-z')g_1(z')
\end{equation}
and can be solved using standard techniques based on the factorization of the
Fourier transformed kernel $G_1(\omega)$
\begin{equation}
  [1-G_1(\omega)]^{-1} = G^{+} (\omega) G^{-} (\omega)\ ,
  \quad \lim_{\omega\to\infty}G^{\pm}(\omega)=1
  \label{WHfac}
\end{equation}
into functions $G^{\pm}(\omega)$ which are analytic for $Im(\omega)>0$ ($<0$),
respectively.  For the $t$--$J$ model these techniques have been applied
before\cite{SchltJ,Essl96} and the factorization of the kernel
is known to be
\begin{equation}
  G^{-}(\omega)=G^{+}(-\omega)=
  \frac{\sqrt{2\pi}}{\Gamma (\frac{1}{2} + i \frac{\omega}{2\pi})}
  \left(\frac{i\omega}{2\pi e}\right)^{\frac{i\omega}{2\pi}}.
  \label{gpgm}
\end{equation}
In Eqs.~(\ref{g1}) and (\ref{gn}) for $n=1$ and $2$ three different driving
terms $g_0$ need to be considered:\newline
(a) The case $g_0(z)=G_0(A+z)$ already appears in the calculation of bulk
quantities such as the dressed energies for the $t$--$J$ model.  Following
Refs.~\onlinecite{SchltJ,Essl96} we find for $g$
\begin{equation}
  g^+_a (\omega) = i G^+ (\omega)  G^- (-i\pi)
  \frac{e^{-\pi A}}{\omega + i\pi}.
  \label{gapS}
\end{equation}
Using the explicit expressions (\ref{gpgm}) we obtain
\begin{equation}
  \int_0^{\infty}\mathrm{d}z\  g_a (z) = g_a^+ (\omega=0) = 
  \sqrt{\frac{2}{e\pi}}e^{-\pi A},
\label{WH_ga}
\end{equation}
which will be necessary to compute the impurity's magnetization.

(b) The second type of driving term, according to (\ref{dtrhosH0}), is $g_0(z)
= G_{\beta}(z+A) + G_{-\beta}(z+A)$.  The analysis of the WH equation is
completely analogous to the first case and we find
\begin{equation}
  g_b^+ (\omega) = 2i G^+ (\omega)G^- (-i\pi)\frac{e^{-\pi A}}{\omega +
    i\pi}\cos ({\pi\beta \over 2}) = 2\cos({\pi\beta \over 2}) g_a^+(\omega).
  \label{gbpS}
\end{equation}

(c) Finally, the driving term in the equation (\ref{gn}) for the sub-leading
contribution, $g_2$, is proportional to $g_0(z)=\int_0^{\infty}\mathrm{d}z'\
G_1 (2A+z+z') g_a (z')$.  Following Ref.~\onlinecite{Essl96}, we perform a
Laplace tansform of $g_0(z)$ to obtain:
\begin{equation}
  g_0(z)\simeq \frac{1}{4\pi} \int_0^{\infty}\mathrm{d}x\ e^{-2Ax} 
  e^{-|z|x} g_a^+ (ix)
  \left(x + \ldots \right),
\end{equation}
where we have used the asymptotic expansion of the function $G_1(z)\sim 1/4\pi
z^2 + {\cal O}(z^{-4})$.  Now the solution of the Wiener--Hopf equation is
given by
\begin{equation}
  g_c^+ (\omega) \simeq G^+ (\omega)\frac{i}{4\pi}\int_0^{\infty}\mathrm{d}x\  
  e^{-2Ax}\left( x+\ldots\right)
  \frac{G^+ (ix)g_a^+ (ix)}{\omega + ix}
  \label{g2p}
\end{equation}
The presence of the rapidly decaying factor $\exp(-2Ax)$ (remember that $A \gg
1$ at small field) in the integrand suggests the following expansion around
$x=0$
\begin{equation}
  g_a^+ (ix)G^+ (ix) \left( x+\ldots\right) \sim
  2 \frac{e^{-\pi A}}{\sqrt{e\pi}}x + {\cal O}(x^2).
\end{equation}
From this expression we obtain
\begin{equation}
  \int_0^{\infty}\mathrm{d}z\ g_c (z) = g_c^+ (\omega=0) = 
  \sqrt{\frac{2}{e\pi}}\,\frac{e^{-\pi A}}{4\pi A}
  +{\cal O}\left( \frac{1}{A^2}\right).
\end{equation}

\section{Expressions for the susceptibilities}
\label{appDC}
For the numerical analysis of the susceptibilities in a finite magnetic field
it is convenient to use their expressions in terms of the so called dressed
charge matrix\cite{FrKo90,BarX92,HUBBARD}.  This quantity is as the
solution of the Bethe ansatz integral equations
\begin{equation}
  \left(\begin{array}{cc} \xi_{ss}(\lambda) & \xi_{cs}(\lambda)\\
                          \xi_{sc}(\vartheta) & \xi_{cc}(\vartheta)
 \end{array}\right) =  \left(\begin{array}{cc} 1 & 0\\
                          0 & 1 \end{array}\right) +
\left( \begin{array}{cc} -\int_{-A}^A a_2 & \int_{-B}^B a_1 \\ 
         \int_{-A}^A a_1 & 0\end{array} \right)*
\left(\begin{array}{cc} \xi_{ss}(\lambda) & \xi_{cs}(\lambda)\\
                          \xi_{sc}(\vartheta) & \xi_{cc}(\vartheta)
 \end{array}\right)
\label{igldc}
\end{equation}

The magnetic susceptibility at fixed $\mu$ is obtained by first taking the
derivatives of the magnetization as obtained from (\ref{eqfix}) with respect
to the boundaries of integration $A$ and $B$.  Then, starting from the
conditions $\varepsilon_s(A)=0 =\varepsilon_c(B)$, one obtains two equations
for their derivatives $\partial A/\partial H$ and $\partial B/\partial H$
which are then solved in terms of $\xi$ at the boundaries of integration
\begin{equation}
\left(\begin{array}{cc} Z_{ss} & Z_{cs}\\
                        Z_{sc} & Z_{cc}
 \end{array}\right) =
  \left(\begin{array}{cc} \xi_{ss}(A) & \xi_{cs}(A)\\
                          \xi_{sc}(B) & \xi_{cc}(B)
 \end{array}\right)\ .
\end{equation}
This approach leads to the following expression for the bulk magnetic
susceptibility at fixed chemical potential\cite{BarX92}
\begin{equation}
  \left.\chi_{bulk}\right|_\mu = \frac{1}{4\pi}\left(
             \frac{(Z_{cc}-2Z_{sc})^2}{v_c} +
             \frac{(Z_{cs}-2Z_{ss})^2}{v_s}\right),
\end{equation}
where
and $v_s=\varepsilon_s'(A)/\pi\rho_s(A)$, $v_c=\varepsilon_c'(B)/\pi\rho_c(B)$
are the spinon and holon Fermi velocities, respectively.  The impurity
contribution to the magnetic susceptibility is obtained analogously starting
from (\ref{Mimp}) as\cite{BEFimp}
\begin{equation}
  \left.\chi_{imp}\right|_\mu = \frac{1}{4\pi}\left(
             \frac{(Z_{cc}-2Z_{sc})^2f_c}{v_c} +
             \frac{(Z_{cs}-2Z_{ss})^2f_s}{v_s}\right),
\end{equation}
with $f_s = \rho^{(R)}_s(A)/\rho_s(A)$ and $f_c = \rho^{(R)}_c(B)/\rho_c(B)$.

In this paper we work at fixed (bulk) hole concentration $\delta=\int_{-B}^B
\mathrm{d}\vartheta\ \rho_{\infty,c}(\vartheta)$.  Using
$\partial\delta/\partial H$ together with the expression of the chemical
potential entering Eqs.~(\ref{de}) in terms of the independent variables $H$
and $\delta$ the bulk susceptibility is found to be (see also
Refs.~\onlinecite{FrKo90,HUBBARD})
\begin{equation}
  \left.\chi_{bulk}\right|_\delta = \frac{(\det Z)^2}{\pi}\,
           \frac{1}{v_c Z_{cs}^2 +v_s Z_{cc}^2}\ .
\end{equation}
Again, it is straightforward to compute the impurity contribution to the
susceptibility within this approach giving
\begin{equation}
  \left.\chi_{imp}\right|_\delta = \frac{\det Z}{\pi}\,
           \frac{1}{v_c Z_{cs}^2 +v_s Z_{cc}^2}
           \left\{
              f_s Z_{cc}(Z_{ss}-\frac{1}{2}Z_{cs})
             -f_c Z_{cs}(Z_{sc}-\frac{1}{2}Z_{cc})\ 
           \right\}\ .
\end{equation}

To measure the valence fluctuations on the impurity site one has to consider
the charge compressibility.  Again, the bulk and impurity contributions to
$\kappa=\partial N_e/\partial \mu$ are conveniently expressed in terms of the
dressed charge matrix\cite{FrKo90,BarX92,BEFimp,HUBBARD}
\begin{equation}
\begin{aligned}
  \left.\kappa_{bulk}\right|_H =&
   \frac{{Z_{cc}}^2}{\pi v_c} + \frac{{Z_{cs}}^2}{\pi v_s}\ ,\\
  \left.\kappa_{imp}\right|_H =&
   \frac{{Z_{cc}}^2}{\pi v_c}\,f_c + \frac{{Z_{cs}}^2}{\pi v_s}\,f_s\ .
\end{aligned}
\end{equation}

\newpage

\newpage
%

%

\begin{figure}[ht]
(a) \includegraphics[width=0.7\textwidth]{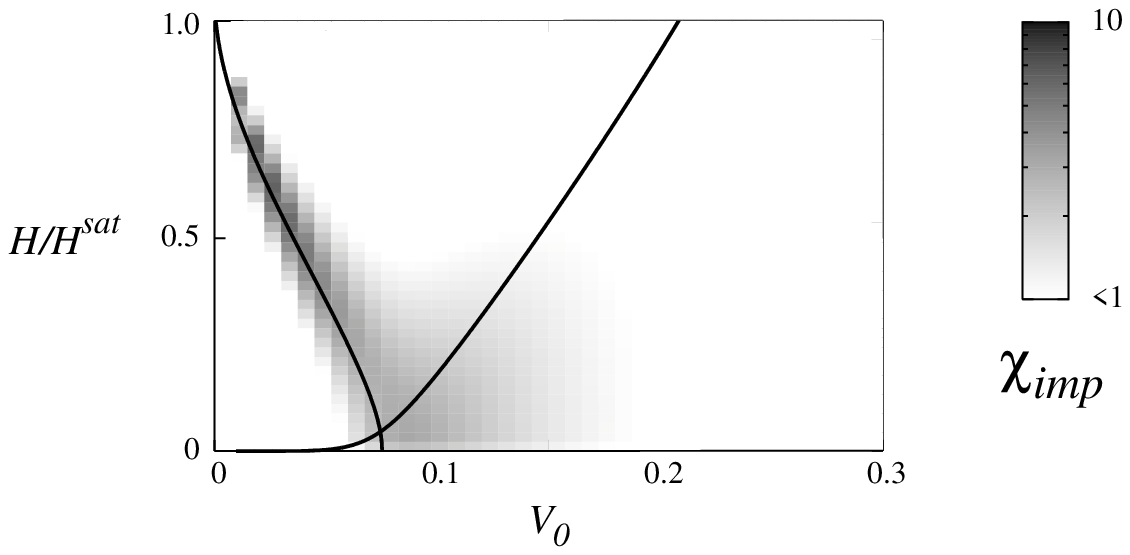}\newline
(b) \includegraphics[width=0.7\textwidth]{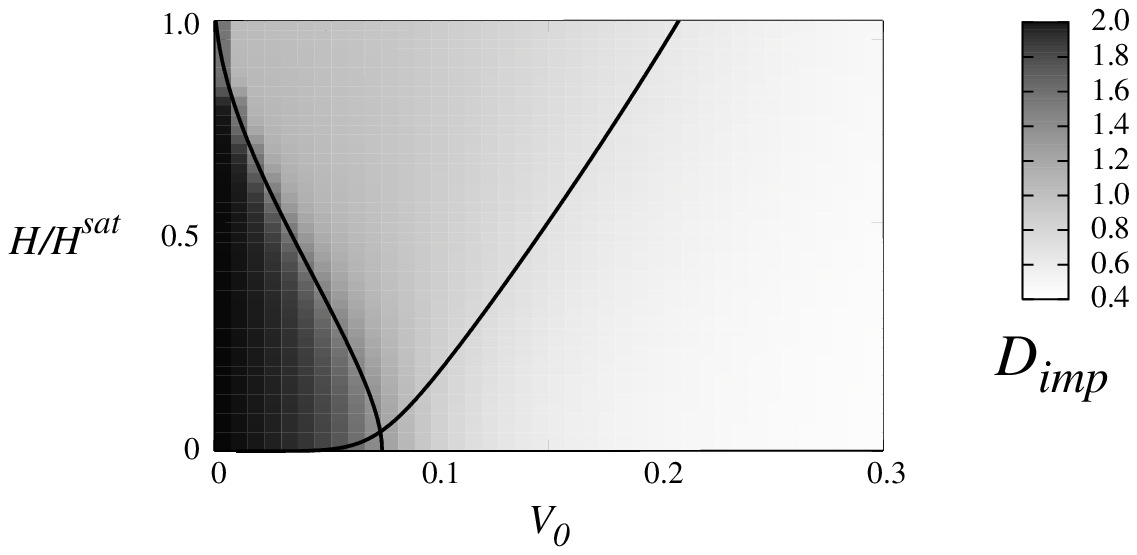}\newline
(c) \includegraphics[width=0.7\textwidth]{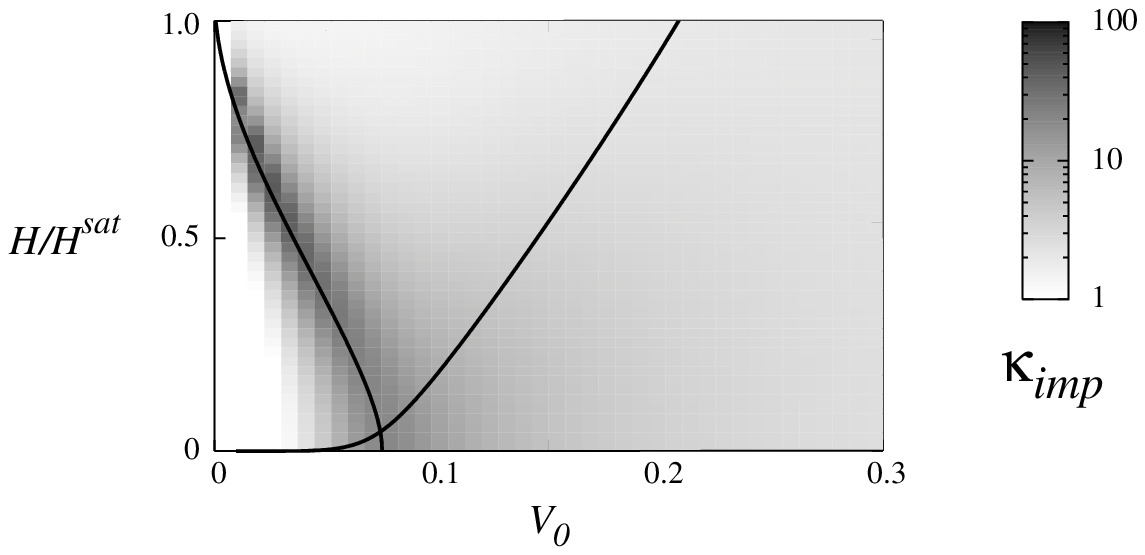}\newline
  \caption{
    \label{fig:IMP}
    (a) Susceptibility $\chi_{imp}$, (b) electronic occupation $D_{imp}$, and
    (c) charge compressibility $\kappa_{imp}$ (normalized to its bulk value)
    of the impurity as a function of the hybridization and magnetic field for
    $\delta=0.8$ and $\alpha=1$ (phase boundaries from Fig.~\ref{fig:Hscales}
    are superimposed).  Note the logarithmic scale used for the shading of the
    susceptibilities.}
  \end{figure}

\end{document}